\documentstyle[12pt,epsf]{article}
\newcommand{\sect}[1]{\section{#1}\setcounter{equation}{0}}
\newcommand{\roughly}[1]{\raise.3ex\hbox{$#1$\kern-.75em
\lower1ex\hbox{$\sim$}}}

\def\centeron#1#2{{\setbox0=\hbox{#1}\setbox1=\hbox{#2}\ifdim
\wd1>\wd0\kern.5\wd1\kern-.5\wd0\fi
\copy0\kern-.5\wd0\kern-.5\wd1\copy1\ifdim\wd0>\wd1
\kern.5\wd0\kern-.5\wd1\fi}}
\def\ltap{\;\centeron{\raise.35ex\hbox{$<$}}{\lower.65ex\hbox{$\sim$}}\;}
\def\gtap{\;\centeron{\raise.35ex\hbox{$>$}}{\lower.65ex\hbox{$\sim$}}\;}
\def\gsim{\mathrel{\gtap}}
\def\lsim{\mathrel{\ltap}}

\def\beq{\begin{equation}}
\def\eq{\end{equation}}
\def\eeq{\end{equation}}
\begin{document}
\bigskip
\hskip 5in\vbox{\baselineskip12pt
\hbox{NSF-ITP-01-62}
\hbox{SU-ITP-01-30}
\hbox{hep-ph/0106219}}
\bigskip\bigskip\bigskip

\centerline{\Large\bf High Energy Colliders as Black Hole Factories:}
\vspace*{.15in}
\centerline{\Large\bf The End of Short Distance Physics}
\bigskip
\bigskip
\bigskip
\centerline{{\bf Steven B. Giddings}$^{1,2}$ and {\bf Scott Thomas}$^{2,3}$}
\bigskip
\bigskip
\centerline{$^{1}$Department of Physics, University of
California, Santa Barbara, CA\ 93106}
\smallskip
\centerline{$^{2}$Institute for Theoretical Physics, University of
California, Santa Barbara, CA\ 93106-4030}
\smallskip
\centerline{$^{3}$Department of Physics, Stanford University,
Stanford,
CA\ 94305}
\bigskip

\begin{abstract}
If the fundamental Planck scale is of order a TeV, as the case in some
extra-dimensions scenarios, future hadron colliders such as the Large
Hadron Collider will be black hole factories.   
The non-perturbative process of black hole formation and
decay by Hawking evaporation gives rise to spectacular events with up to
many dozens of relatively hard jets and leptons, with a
characteristic ratio of hadronic to leptonic activity of roughly
5:1.  The total transverse energy of such events is typically a sizeable
fraction of the beam energy.  
Perturbative hard scattering processes at energies well above the 
Planck scale are cloaked behind a horizon, thus limiting the ability 
to probe short distances. 
The high energy black
hole cross section {\it grows} with energy at a rate determined by the
dimensionality and geometry of the extra dimensions.
This dependence therefore probes the extra dimensions at 
distances larger than the Planck scale. 
\end{abstract}
\newpage

\baselineskip=17pt


An outstanding problem in physics is
the ratio between the four dimensional Planck scale
$G_N^{-1/2} \sim 10^{19}$ GeV and the
electroweak scale $G_F^{-1/2} \sim 100$ GeV.
Scenarios have emerged recently that
address this hierarchy within the context of the old idea that
the Standard Model is confined to a brane in a higher dimensional
space.
In this case the fundamental Planck scale -- the energy
at which gravitational interactions become strong --
can be as low as the TeV scale.
This raises the exciting possibility that
future high energy colliders can directly probe strongly
coupled gravitational physics.
In this paper we investigate the dramatic TeV scale gravity 
signatures associated with 
black hole production and evaporation at high energy colliders.

A description of scattering processes at center of mass energies
of order the fundamental
Planck scale requires a full theory of quantum gravity.
However, as we'll discuss, 
scattering at energies well above the Planck scale
is believed to be described in any theory by
semi-classical general relativity.
Probably the most interesting phenomenon in this regime is the
production and subsequent evaporation of black holes.
Black hole intermediate states are in fact believed to
dominate $s$-channel scattering
at super-Planckian energies.  Indeed, the number of such
non-perturbative states
grows faster than that of any perturbative state;
for example, the number of black hole states in $D$
space-time dimensions grows with mass like
${\rm exp}(M^{(D-2)/(D-3)})$ while the number of perturbative string states
grows like ${\rm exp}(M)$.
The importance of black holes is also apparent in a geometric
picture -- scattering impact parameters which are smaller than the
Schwarzschild radius associated with the center of mass energy
result in black hole formation.
In the high energy limit this classical non-perturbative process
leads to a cross section which
{\it grows} with energy much faster than that associated with
any known perturbative local physics.  Some of 
these features of high energy scattering have 
been discussed  previously in
\cite{Banks:1999gd}.

The observability of black hole production at future colliders 
of course depends on the value of the fundamental Planck scale. 
The present bound on the Planck scale in a brane-world
scenario with $D=10$ space-time dimensions arising from 
missing energy signatures at the Tevatron Run I and LEP II
associated with perturbative graviton production 
is just under $M_p \gsim 800$ GeV \cite{missE,missErev}
in the standard normalization given below. 
As detailed below, if the fundamental Planck scale is of order a 
TeV, the black hole cross section at the 
Large Hadron Collider (LHC) is large enough 
to qualify the LHC as a black hole factory. 
This opens up the possibility of detailed experimental study 
of black hole production and decay,  
in addition to perturbative quantum gravity processes. 

The specific signatures associated with black holes produced in high 
energy collisions depend on the decay products. 
The decay of an excited spinning black hole state 
proceeds through several stages.
The initial configuration looses hair associated with multipole moments
in a {\it balding} phase by the emission of classical gravitational
and gauge radiation.
Gauge charges inherited from the initial state partons 
are discharged by Schwinger emission. 
After this transient phase, the subsequent spinning black hole
evaporates by semi-classical Hawking
radiation in two phases:  a brief {\it spin-down} phase in which
angular momentum is shed, and a longer {\it Schwarzschild} phase.
The evaporation phases give rise to a large number of quanta 
characteristic of the total entropy of the initial black hole. 
At present it is not possible to describe quantitatively the
end point of Hawking evaporation
but a reasonable expectation is that
when the black hole mass decreases to the fundamental Planck scale
it enters a {\it Planck} phase in which final decay takes place
by the emission of a few Planck scale quanta.
Most of the black hole decay products are Standard Model 
quanta emitted on the brane \cite{Emparan:2000rs}
and are therefore visible experimentally.

The large number of visible quanta emitted in the 
decay of a black hole 
gives rise to the very distinctive signature of large multiplicity 
events with large total transverse energy, as described in detail 
below. 
The observation of such events with a parton level cross section 
which grows with energy would be a smoking gun for black hole 
production. 

As discussed below, black hole production cross sections rise with energy
whereas cross sections from other conventional hard 
perturbative processes
should fall.  Correspondingly, as the energy grows, the range of impact
parameters for which black holes are formed  grows, and other hard
processes will be cloaked and invisible due to the formation of the event
horizon\cite{Banks:1999gd}.
This means that the era of black hole formation represents the
end of experimental short distance physics.  Nonetheless, scattering at
higher energies can remain interesting, as it can begin to reveal the
structure of the extra dimensions on scales large as compared to the Planck
scale through features such as the energy dependence of the cross section.

In the next section two classes of brane-world scenarios
with flat and warped geometries for the extra dimensions 
are reviewed. 
The relevant black hole properties of Schwarzschild radius, 
temperature, and entropy are presented for the two scenarios. 
The question of applicability of a black hole description for a 
generic high energy state produced in a collision 
is also addressed. 
In section \ref{sec:production} the cross section 
for black hole production is discussed. 
Production rates for the LHC are estimated and shown to be sizeable
for a Planck scale of a TeV. 
The initial balding phase of black hole decay in which multipole moment 
hair is shed by gauge and gravitational radiation is also described. 
Section \ref{sec:evap} describes the spin-down and Schwarzschild eras
of Hawking evaporation. 
The total number and energy of quanta emitted in this phase is shown to be 
characteristic of the initial black hole entropy and Hawking 
temperature respectively. 
The spectacular signatures resulting from black hole production 
and decay which could be observed at the LHC if the 
fundamental Planck scale is a TeV are described in section \ref{sec:sig}.
The final section includes a discussion of the dependence of the
high energy black hole cross section on the dimensionality and 
geometry of the extra dimensions as a manifestation of the 
infrared--ultraviolet connection in a theory of gravity, and 
implications for the future of 
experimental short-distance physics are also discussed. 

%
%
%
%
%
%
%
%
%
%

\sect{TeV Scale Gravity and Black Holes}
\label{sec:grav}

One scenario  
for realizing TeV scale gravity is a brane-world
in which the Standard Model matter and gauge degrees of freedom
reside on a 3-brane within a flat compact space of volume 
$V_{D-4}$ \cite{Arkani-Hamed:1998rs,Antoniadis:1998ig}.
Gravity propagates in both the compact and non-compact dimensions.
The Einstein action
\begin{equation}
S_E={1\over 8\pi G} \int d^D x \sqrt{-g} ~{1 \over 2} {\cal R} 
\label{Eact}
\end{equation}
implies
the relation between  
the four-dimensional and $D$-dimensional Newton's constants:
\beq
G_N = {G_D \over  V_{D-4} }\ .
\eq
The reduced fundamental Planck scale in this case is generally related to 
the $D$ dimensional Newton's constant 
in the phenomenology literature\footnote{Here we use the conventions
of \cite{missErev}, which differs from the conventions for the Planck
mass $M_D$ in the first
reference of \cite{missE} as $M_p = 2^{1/(D-2)} M_D$.}  as
\beq
M_p^{D-2}={(2 \pi)^{D-4} \over 4 \pi G_D } \ .
\label{Mpnorm}
\eq
The fundamental Planck scale can in principle be experimentally accessible 
in high energy collisions if
$V_{D-4}$ is large in fundamental units.
For example, $M_p \sim $ TeV for $D=10$ with
$V_6 \sim {\rm fm}^6$ \cite{Arkani-Hamed:1998rs}.
Although the total volume of the compact space must be large 
in fundamental units, the radii of some of the extra dimensions, 
$R_c$, can in principle be not much larger than the fundamental scale. 
We refer to this as the ``flat scenario.''

Another scenario for realizing TeV scale gravity 
arises from properties of warped extra-dimensional geometries pointed out
in \cite{Randall:1999ee}.
Examples of string theory solutions that generate
a hierarchy this way
were recently exhibited in 
\cite{Giddings:2001yu}.
A warped geometry is described by a
metric of the form
\begin{equation}
ds^2=e^{2A(y)} dx_4^2 + g_{mn}(y) dy^m dy^n\ .
\label{warpedmet}
\end{equation}
Here $dx_4^2= \eta_{\mu\nu} dx^\mu dx^\nu$ is
the standard four-dimensional Minkowski line element, and
the coordinates $y$ parameterize the extra dimensions of spacetime,
with metric $g_{mn}$.  
The function $e^A$ is the warp factor, and leads to
scales for four-dimensional physics that depend on location  
within the extra dimensions.  Gravity propagates in both the
compact and non-compact directions. 
As we see from the action (\ref{Eact}),
in this case 
the four-dimensional Newton's constant is related to the the $D$ 
dimensional one by 
\begin{equation}
G_N^{-1} = G_D^{-1} \int d^{D-4}y  \sqrt{g}~ e^{2A}\ .
\end{equation}
The Standard Model confined to a brane at $y=y_0$ within such a
geometry will have a Lagrangian of the form
\begin{equation}
S_{SM}= \int d^4 x e^{4 A(y_0)} {\cal L}(e^{2A(y_0)}
\eta_{\mu\nu}, \psi_i, m_i) \label{SMlag}
\end{equation}
in which the metric $\eta_{\mu\nu}$ appears accompanied by
factors of $e^{2A(y_0)}$. 
Here $\psi_i$ denote the matter fields, and $m_i$
are mass parameters which naturally take values 
of order the conventional Planck scale $M_p$. 
By
rescaling the kinetic terms in (\ref{SMlag}) to canonical forms, 
the measured four-dimensional masses all take values of order 
$e^{A(y_0)} M_p$; alternatively one may use the redundancy under
$A\rightarrow A+\lambda$ and $x\rightarrow e^{-\lambda}$ in the metric
(\ref{warpedmet}) to choose units in which the masses are ${\cal O}(M_p)$.  
If the warp factor $e^A$ is small in the vicinity of the 
Standard Model brane, particle masses can take TeV
values (or, in the
rescaled units, the extra dimensions are large as in the flat scenario),
thereby giving rise to a large hierarchy between the TeV and conventional 
Planck scales. 
Conversely, high energy scattering processes on the brane 
at apparent energy scales of order TeV actually probe energies 
approaching the fundamental Planck scale $M_p$, and can 
probe strong gravitational effects including 
black hole formation \cite{Giddings:2000ay}.
We will refer to this as the
``warped scenario.''

In either scenario the Planck scale threshold for strong gravity
effects can be in the TeV range, and collider physics at
such energies may reveal a wealth of fascinating and new physics.  If this
is the case, description of physics in this regime requires a quantum
theory of gravity, such as string theory, which would predict many new
effects. However, a generic effect in any theory of quantum gravity is the
formation of black holes.  While a quantitative 
understanding black holes with masses of order the Planck
scale is quite difficult, for masses well above this scale 
black holes exhibit many features well described by semi-classical
physics.  
And, as discussed below, it is possible that 
black holes not too much heavier than 
the fundamental Planck scale may be produced at future colliders. 

In order to discuss black hole production and evaporation
in the laboratory we 
therefore consider black holes with masses
$M \gsim {\rm (few)}M_p$
where features of the semi-classical analysis are expected to begin to be
valid.  Several simplifying assumptions are appropriate. 
To begin with, the
brane on which the Standard Model lives will have a gravitational field
which should be accounted for in solving Einstein's equations.  While some
features of such solutions were discussed in
\cite{chr,Emparan:2000wa,Garriga:2000yh,Giddings:2000mu}, we will assume that
the only effect of the brane field is to bind the black hole to the brane,
and that otherwise the black hole may be treated as an isolated black hole
in the extra dimensions; this is the ``probe brane approximation.''   
Secondly, initially we assume that
the black hole can be treated as a solution
in $D$ flat space-time dimensions.
This assumption will be
valid in the large-dimensions scenario at distance small compared 
with any radii, $r\ll R_c$, and in the
warped scenario at distances small as compared to the curvature scale of
the geometry associated with the extra 
dimensions, which we also denote as $R_c$.  Finally,
string theory has a number of other fields such as the dilaton; we will 
assume that these are fixed and do not play an important role in the
relevant black hole solutions. We will also argue that gauge 
charges don't have a big effect on the black hole solutions, but will see
that spin of the black
holes is important.  For an earlier discussion of some properties of black
holes in these approximations, see 
\cite{Argyres:1998qn}. 

Black holes relevant to experimental investigation in the 
laboratory are therefore 
neutral but spinning solutions of the $D$ dimensional
Einstein action (\ref{Eact}).
These are the higher-dimensional Kerr solutions discussed in
\cite{Myers:1986un}.
While we will not rewrite such solutions explicitly, let us recall some of
their salient features.  In general these have $[(D-1)/2]$ angular momentum
parameters $J_i$, but as argued below only a single angular momentum
parameter $J$ parameterizing the four-dimensional spin is relevant.
The horizon radius is given by 
\begin{equation}
r_h^{D-5}\left(r_h^2+{(D-2)^2J^2\over 4M^2}
\right) = 
{16 \pi G_D M \over (D-2) \Omega_{D-2} }
\,\, \buildrel J \rightarrow 0 \over\longrightarrow \,\,
r_h=\left[{ 4(2\pi)^{D-4} ~ M \over
(D-2) \Omega_{D-2} ~M_p^{D-2}}\right]^{1/ (D-3)}\ ,
\label{rsch}
\end{equation}
where 
$$
\Omega_{D-2}= { 2\pi^{(D-1)/2} \over  \Gamma({D-1 \over 2})}
$$
is the area of a unit $D-2$ sphere.
Note that the horizon size grows with mass like a power
that depends on the space-time dimension, 
$r_h \propto M^{1/(D-3)}$. 
For later convience it is 
useful to introduce a dimensionless rotation parameter,
\begin{equation}
a_* = {(D-2)J\over 2M r_h}\ .
\end{equation}
According to Hawking such black holes are unstable 
semi-classically \cite{Hawking:1975sw}, and decay into
an approximately thermal spectrum of particles, 
with a temperature given by
\begin{equation}
T_H = {D-3 + (D-5)a_*^2 \over 4\pi r_h(1 + a_*^2) } \,\,
\buildrel J\rightarrow 0 \over\longrightarrow \,\,
{D-3\over 4\pi r_h}\ .
\label{Thawk}
\end{equation}
Correspondingly, the black holes have entropy 
\begin{equation}
S_{bh} = {M\over (D-2) T_H}\left(D-3-{2a_*^2 \over  1 +a_*^2}\right)\,\,
\buildrel J\rightarrow 0 \over\longrightarrow\,\,
{r_h^{D-2} \Omega_{D-2} \over 4G_D} \ .
\label{Sbh}
\end{equation}

For a horizon radius $r_h\sim R_c$ the higher dimensional 
Kerr solutions are no longer
a valid description.  First consider the flat scenario.  
For horizon sizes larger than some radii, 
$r_h \gsim R_c$, the relevant solution is instead a 
lower dimensional black hole solution
extended uniformly over the extra dimensions with small radii.  
The mass dependence of the horizon size
for black holes larger than these radii would be that for the 
lower dimensional space-time. 
A somewhat similar
phenomenon is expected in the warped case.  
A heuristic argument for the form of the black hole solutions 
in this case follows from
the linearized approximation.  There we expect the metric to be of the form
\begin{equation}
ds^2 \simeq -\left(1 - {16 \pi G_D~M \over (D-2) \Omega_{D-2}r^{D-3}}
   + V(\rho)\right)dt^2
    +\cdots \label{gtt}
\end{equation}
where $\rho$ is the distance transverse to the brane
and $V(\rho)$ is an effective gravitational potential
associated with the warping and
curvature in the extra dimensions.
$V(\rho)$ grows with increasing $\rho$ and
by definition becomes important for $\rho\sim
R_c$. 
The effective potential 
slows the growth of the horizon into the extra dimensions 
(determined by the vanishing
of (\ref{gtt})) for increasing black hole mass. 
We expect this to be qualitatively similar to the 
inability of the horizon to
grow transversely past $R_c$ in the large-dimensions case.  
The horizon can, however, grow 
in the flat four-dimensional directions with increasing mass. 
This implies a crossover of the mass dependence of the horizon size
from that for flat $D$ dimensional space-time given above for 
$r_h \lsim R_c$ to a modified functional dependence on M for $r \gsim R_c$,
with a corresponding modification of the temperature and entropy formulas
(\ref{Thawk}), (\ref{Sbh}).
Note that, in either scenario, there may be multiple thresholds where
these formulas change, caused either by the existence of different radii
for the extra dimensions, or by different curvature scales
encountered in warped compactifications.

If the fundamental Planck scale is of order a TeV, 
center of mass energies not too much larger than the fundamental 
Planck scale will be available for producing black holes. 
It is therefore important to address the applicability of 
the description of a generic massive state as a semi-classical black hole
in this mass range. 
For masses of order the fundamental Planck scale there is no 
control over 
quantum gravity effects which are likely to invalidate the 
semi-classical and statistical thermodynamic pictures. 
A precise criterion for the quantum corrections to be small 
is hard to formulate and would ideally be studied in
the context of a quantum theory for gravity.  
One measure of the quantum corrections to a semi-classical 
black hole 
is the change in Hawking temperature per particle emission.
A necessary condition for the quantum back-reaction on the 
black hole geometry to be small, 
and for a quasi-static classical description of the metric to be
good, is that this quantity is small compared with the 
Hawking temperature\cite{Preskill:1991tb}
\begin{equation}
T_H \left|{\partial T_H\over \partial M}\right| \ll T_H\ .
\label{dT}
\end{equation}
A second criterion based on the statistical thermodynamic 
interpretation of a black hole is 
that the fluctuations in a micro-canonical description be small. 
Since the number of degrees of freedom in an ensemble 
describing a black hole is roughly the entropy, small 
statistical fluctuations require $1/\sqrt{S_{bh}} \ll 1$. 
These criteria are
related, 
since from (\ref{rsch}), (\ref{Thawk}), and (\ref{Sbh})
\begin{equation}
{\partial M \over \partial T_H} = (2-D) S_{bh}\ .
\end{equation}
The first criteria is then equivalent to $S_{bh} \gg 1$, 
while the second more stringent statistical one is 
$\sqrt{S_{bh}} \gg 1$.
Numerically for $D=10$ the entropy is 
$S \simeq 4 \left({M / M_p}\right)^{8/7}$. 
With a fundamental Planck scale of $M_p \simeq 1$ TeV
the entropy of a 5 TeV mass black hole is $S_{bh}(5M_p) \simeq 27$, 
while for a 10 TeV mass black hole 
$S_{bh}(10M_p) \simeq 59$.
Since a specific numerical criterion on the mass at which 
the black hole description becomes valid is subjective, we 
consider both masses given above for the cross section estimates below. 
For a smaller number of space-time dimensions the growth of 
the entropy with mass is more rapid, implying a slightly 
lower mass for which the black hole description should be 
valid. 

Another necessary requirement for the validity of 
a description of black hole production and decay is that the 
lifetime determined by the Hawking evaporation process be 
sufficiently larger than the mass, $\tau M \gg 1$. 
In this case a black hole is a well defined resonance, 
and may be thought of as an intermediate state in the $s$-channel.
The parametric dependences and numerical estimates for black 
hole lifetimes are presented in section \ref{sec:evap}.
Numerically, for $D=10$ and $M_p=1$ TeV the lifetime of a 
5 TeV mass black hole is estimated very roughly to be 
$\tau(5M_p) \sim 10 M^{-1}$, while for a 10 TeV mass black hole 
$\tau(10M_p) \sim 12 M^{-1}$.

In a weakly coupled string theory 
another requirement for the 
validity of a semi-classical black hole description comes
from possible stringy corrections to the classical geometry. 
Black holes with 
horizon size
less than the string length would receive large corrections.
Conversely, a generic string state larger than the string scale 
is a semi-classical black hole \cite{hp}. 
In typical models where the 
string and Planck scales are not widely 
separated the above conditions on the 
validity of a black hole description of a generic massive state 
produced in a high energy collision are not significantly modified. 
For example, in $D=10$, the string and Planck scales are related 
by $L_{\rm string} \sim 1/(g^{1/4} M_p)$, with $g$ the string coupling. 
In cases where the Planck scale is significantly higher than the string
scale, there is an intermediate regime where excited string states would be
produced; for some discussion of the phenomenology of such a scenario see 
\cite{Cullen:2000ef}.



\sect{Black Hole Production and Balding}
\label{sec:production}

Particle scattering at super-Planckian energies is dominated
in the $s$-channel by black hole production.
In this limit the eikonal approximation for the initial state becomes
valid and is described by a metric which
contains a pair of Aichelburg-Sexl shock waves \cite{shock}
with impact parameter $b$.
A classical picture of black hole formation in this
metric should capture the essential features
of the scattering process in the high energy limit.
For an impact parameter $b \lsim r_h$, where $r_h$ is the
Schwarzschild radius associated with the center of mass
energy $\sqrt{s}$, the incident relativistic particles
pass within the event horizon. 
Formation of the 
event horizon should occur before the particles come in causal contact
and be a classical process. 
Once inside the event horizon, no matter how violent and non-linear 
the subsequent collision, formation of an excited black hole state
results. 
The process of scattering two particles, $i$ and $j$, confined
on a 3-brane, to form a $D$-dimensional
black hole as shown in Fig. 1 may then 
be modeled by a scattering amplitude described by an
absorptive black disk with area $\pi r_h^2$.
This gives a cross section 
\beq
 \sigma_{ij \rightarrow bh}(s) ={F}(s) \pi r_{h}^2 =
{F}(s)
\pi \left( {4(2 \pi)^{D-4}~ \sqrt{s}
   \over (D-2) \Omega_{D-2} ~ M_p^{D-2} }
   \right)^{2/({D-3})}
   \label{bhcross}
\eq
where ${F}(s)$ is a dimensionless order one form factor coefficient,  
and where a black hole is by definition any matter or energy
trapped behind the event horizon formed by the collision. 
Even though the  process of black hole formation is a highly non-linear,
non-perturbative process, 
formation of the event horizon ensures that the black disk
approximation gives the correct magnitude and parametric dependence
of the amplitude in the high energy limit. 
The precise mass of the black hole formed in a collision depends 
on the the amount of energy and matter which becomes trapped 
behind the event horizon. 
In the high energy limit this in turn depends on the impact parameter
$b$. 
So a range of black hole masses will result for a given center of
mass energy 
\beq
{F}(s) = \int dM~ { d {F}(s,M) \over dM}\ .
\eq
However, since the cross section is dominated geometrically 
by large impact parameters, $b \lsim r_h$, the average black hole
mass should be of order of the $ij$ center of mass energy, 
$\langle M \rangle \lsim \sqrt{s}$. 
The precise order one coefficient ${F}(s)$ 
appearing in the
cross section (\ref{bhcross}), 
as well as the distribution $d {F}(s,M)/dM$,
could be calculated in the high energy
limit numerically
within classical general relativity by evolving the colliding
Aichelburg-Sexel shock waves and integrating over a
range of impact parameters.

\begin{figure}[ht]
\begin{center}
\epsfxsize= 3.95 in
\leavevmode
\epsfbox[0 0 577 164]{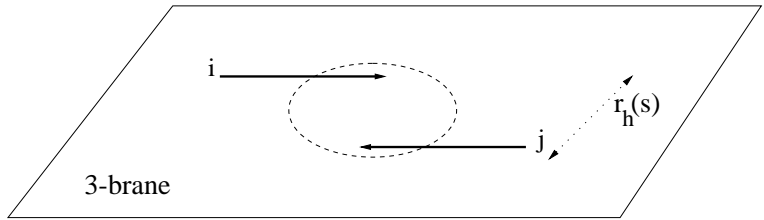}
\end{center}
\caption[f1]{Two partons, $i$ and $j$, form a black hole by 
passing within the event horizon determined by the 
Schwarzschild radius associated with 
the center of mass energy $\sqrt{s}$.}
\label{bhform}
\end{figure}

The cross section for black hole production (\ref{bhcross}) has a number
of interesting and important features.
Since it is a classical non-perturbative gravitational phenomenon, 
it contains no small numbers or coupling constants.
As such, black hole production would not appear at any 
order in perturbation theory. 
Most strikingly, the cross section {\it grows} with
the center of mass energy like a power which depends on the dimensionality
of space-time.
This is connected with the rapid growth of the density of black hole states
at large mass.
It may also be understood as a
manifestation of the infrared--ultraviolet connection
within gravitational theories -- super-Planckian energies correspond to
large rather than short distances.
A power law growth of the cross section with energy does not result
from any known perturbative local physics, and is one of the
most characteristic feature of black hole formation.
Additionally, in the high energy limit in which the
classical picture of formation is valid, a black hole can be formed for any
incident center of mass energy.
The black hole 
may therefore be thought of as a intermediate resonance
with effectively a continuum of states representing the large number of
black hole states.
In this language the intermediate black hole state produced in a given 
collision should not be thought
of as a single massive degree of freedom, but rather as a 
state with a number of degrees of freedom (given by the entropy) 
in approximate statistical thermodynamic equilibrium. 
For black hole masses close to the fundamental Planck scale
a full quantum treatment is necessary to address analogous 
non-perturbative scattering processes. 
However, for center of mass energies for which semi-classical 
black holes are well defined objects a semi-classical  
description of the event horizon should be applicable and 
the geometric cross section (\ref{bhcross})
should therefore provide a reliable estimate. 
As discussed in the previous section, this may be achieved for 
masses not far above the fundamental Planck scale.

The non-perturbative process of black hole formation 
in high energy collisions also has an effect on the amplitudes
for other processes. 
Hard perturbative processes at center of mass energies well 
above the fundamental Planck scale should be 
highly suppressed. 
This may be understood from the statistical thermodynamic 
properties of semi-classical black holes \cite{Banks:1999gd}.
The amplitude for a massive black hole with a significant entropy
to decay to a state with a few very energetic particles is 
Boltzmann suppressed. 
And since a generic state at energies far above the fundamental 
Planck scale is well approximated by a black hole, the high 
energy amplitude
for $2 \rightarrow$ few scattering is Boltzmann suppressed
compared with the Hawking emission final state resulting 
from an intermediate black hole. 
This suppression may also be understood geometrically. 
In the high energy limit an event horizon forms before the particles 
come in causal contact. 
Any hard processes taking place at short distances are therefore 
cloaked behind a horizon and can not lead to final state 
hard quanta which escape the scattering center. 
This feature also has the effect of suppressing initial state 
radiation in the high energy limit. 

The only colliders envisioned which can reach energies well above
the TeV scale, and therefore potentially produce black holes,
are hadron colliders.
In order to obtain the $pp \rightarrow bh$ cross section, the 
parton cross section (\ref{bhcross}) must be convoluted with the
parton distribution functions (as long as the cross section is 
smaller than the geometric area of the proton).
An intermediate resonance produced in a parton collision must
carry the gauge and spin quantum numbers of the initial parton pair.
In the high energy limit, black hole states exist with
gauge and spin quantum numbers corresponding
to any possible
combination of quark or gluon partons within the protons.
The $pp \rightarrow bh$ cross section therefore includes
a sum over {\it all} possible parton pairings
\begin{eqnarray}
\sigma_{pp \rightarrow bh}(\tau_m,s) &=&
  \sum_{ij} \int_{\tau_m}^1 d \tau \int_{\tau}^1 {dx \over x}
    f_i(x) f_j(\tau/x)
         \sigma_{ij \rightarrow bh}(\tau s) \nonumber \\
                 & \equiv & h(\sqrt{\tau_m}) \sigma_{ij \rightarrow bh}(s)
                 \label{partcross}
\end{eqnarray}
where here $\sqrt{s}$ is the collider center of mass energy, 
$x$ is the parton momentum fraction, $\tau=x_i x_j$
is the parton--parton center of mass energy squared fraction, 
$\sqrt{\tau_m s}$ is the minimum center of mass energy for which
the parton cross section into black holes is applicable, 
and the black hole mass is assumed to be $M \simeq \sqrt{\tau s}$. 
The sum over parton distributions $f_i(x)$
includes a factor of two for $i \neq j$.
The sum over all initial parton pairings represents another enhancement
of the high energy black hole cross section 
compared with standard perturbative processes.
This is in addition to the lack
of small couplings and growth with energy.

The momentum scale squared, $Q^2$, 
at which a parton distribution function 
is evaluated is determined by the inverse length scale associated with 
the scattering process. 
For perturbative hard scattering in a local field theory this 
momentum scale is given by the momentum transfer, which in the 
$s$-channel is the parton--parton center of mass energy $Q^2 \sim s$. 
For the non-perturbative process of $s$-channel 
black hole formation in a 
theory of gravity, however, the relevant length scale is 
the Schwarzschild radius rather than the black hole mass, 
$Q^2 \sim r_h^{-2}$. 
This is a consequence of the infrared--ultraviolet properties
of black hole formation -- scattering at 
super-Planckian energies corresponds to large distances. 


The LHC with a center of mass energy of 14 TeV offers the
first opportunity for black hole production if $M_p \sim$ TeV.
Because of the rapid decrease of the parton distributions at
large $x$, the LHC production cross section falls rapidly with
black hole mass for any space-time dimension $D$ even 
though the intrinsic parton--parton cross section grows 
with energy. 
For production of black holes more massive
than 5 TeV at the LHC, with $M_p=1$ TeV and $D=10$,
using the CTEQ5 structure functions \cite{CTEQ5},
the integrated cross section function is
$h(0.36)\simeq 0.02$,
corresponding to a cross section of 
$2.4 \times 10^5$ fb.\footnote{We thank Tom Rizzo for 
cross section estimates.}
This is a very large cross section by new physics
standards and is only a factor of a few smaller 
than that for $pp \rightarrow t \bar{t}$.
With a luminosity of 30 fb$^{-1}$yr$^{-1}$
such a cross section would correspond to a black hole production
rate of 1 Hz, and would qualify the LHC as a black hole factory.
For production of black holes more massive than 10 TeV at the
LHC, with $M_p=1$ TeV and $D=10$, the integrated cross section function is
$h(0.71)\simeq 5 \times 10^{-7}$,
corresponding to a cross section of roughly 10 fb.
Even at these large masses this corresponds to a production rate
of 3 day$^{-1}$.
Black hole production cross sections at the LHC 
for $D=8,10$ with $M_p=1$ TeV and assuming a form factor
coefficient $F(s)=1$ are 
summarized in Table I.$^2$

\begin{table}
\begin{center}
\begin{tabular}{ccc}
 \hline \hline \\ 
 $M$ & $D=8$ & $D=10$  \\ & & \\
 \hline \\
  5 TeV~ & ~$1.6 \times 10^5$ fb~ & ~$2.4 \times 10^5$ fb~ \\
  7 TeV~ & ~$6.1 \times 10^{3}$ fb~ & ~$8.9 \times 10^3$ fb~ \\
  10 TeV~ & ~6.9 fb~ & ~10 fb~ \\ \\
 \hline \\
\end{tabular}
\end{center}
\caption{Large Hadron Collider
$pp \rightarrow bh$ integrated cross sections 
for black holes of mass larger than $M$ with $M_p = 1$ TeV
for $D=8,10$.  
Black holes formed in parton collisions 
are assumed to have mass equal to the parton--parton 
center of mass energy with form factor coefficient $F(s)=1$.} 
\end{table}

With TeV scale gravity, black hole production would become the dominant
process at hadron colliders beyond the LHC.
For example, with $M_p=1$ TeV and $D=10$,
the Very Large Hadron Collider (VLHC)
with 100 TeV center of mass energy and
100 fb$^{-1}$yr$^{-1}$ luminosity would produce black holes
of average mass roughly 10 TeV at a rate of order kHz, and would
produce black holes heavier than 50 TeV at a rate of roughly 0.5 Hz.

The rate of growth of the black hole cross section with 
center of mass energy depends on the black hole density of 
states as a function of mass. 
In a flat background of large uniform volume, 
and for black holes
smaller than the transverse size of the extra dimensions,
it depends on the space-time dimensionality as indicated in 
(\ref{bhcross}). 
In principle, the radii of some of the extra dimensions could 
be comparable to the fundamental Planck scale, 
$R_c \gsim M_p^{-1}$. 
In this case the cross section dependence on center of mass 
energy would increase as a function of energy as the threshold
for producing black holes of size $r_h \sim R_c$ was crossed. 
Alternately, the radius of curvature for a warped background 
can also be comparable to or larger than the Planck scale, 
$R_c \gsim M_p^{-1}$ (see {\it e.g.} \cite{Giddings:2001yu}).
This would also increase the energy dependence of the cross section. 
These dependences illustrate that massive black holes probe 
features of the extra dimensions on scales larger than the Planck 
scale -- another manifestation of the infrared--ultraviolet 
connection in gravitational theories. 



Black holes which are formed in 
high energy collisions have non-vanishing 
angular momentum which is determined by the impact parameter. 
Since the impact parameter is only non-vanishing in directions
along the brane, the angular momentum lies within the brane directions. 
For a given parton-parton center of mass energy a range of 
angular momenta will result for the range of the impact parameters
which lead to a black hole.  
The order one form factor coefficient of the cross section (\ref{bhcross})
therefore in general depends on both the mass and angular momentum 
of the black hole formed in a collision 
\beq
{F}(s) = \int dM~dJ~ { d^2 {F}(s,M,J) \over dM~dJ}
\eq
Since the production cross section is dominated geometrically 
by impact parameters $b \lsim r_h$, 
the black holes will typically be formed
with large angular momentum components in the brane directions,
$\langle J \rangle \sim Mr_h$.
The direction of the spin axis within the 
Standard Model brane is perpendicular to the collision 
axis in the high energy limit.
In the high energy limit, 
the distribution of both masses and angular momenta could 
be calculated numerically within general relativity 
as described above, and are 
presumably correlated through the initial impact parameter. 

Before any radiation of excess energy, the black hole state
will also carry gauge quantum 
numbers of the initial state parton pair. 
In addition, since formation is a violent process
the initial horizon is likely to be very asymmetric.
The excited black hole state then carries
additional hair corresponding to
multipole moments for the distribution of gauge 
charges and energy--momentum within the asymmetric configuration. 

An excited black hole state produced in a collision will shed the
hair associated with the multipole moments during an initial 
transient balding phase. 
In the large mass limit this occurs through classical gauge radiation 
to gauge fields on the brane, and through gravitational radiation. 
The frequency of this radiation, or equivalently the energy 
of emitted quanta, is determined by the frequency of oscillation 
of the multipole moments.
Both the frequency of multipole oscillation and the balding 
time scale are characterized by the 
Schwarzschild radius, $\omega \sim 1/r_h$ and $\tau_b \gsim r_h$. 
The rate of energy loss 
for gauge and gravitational radiation in the balding phase 
can be estimated parametrically. 
Power emitted in gauge radiation should be dominated by 
the dipole mode, $d_i \sim g r_h$, where $g$ is a gauge coupling 
constant. 
Ignoring any prefactors the parametric dependence of such 
dipole power loss is 
\begin{equation}
P_{\rm gauge}\sim {\alpha\over r_h^2}\ ,
\end{equation}
where $\alpha$ is a fine structure constant. 
Power emitted in gravitational radiation should be dominated 
by the energy-momentum quadrupole moment, 
$Q_{ij} \sim M r_h^2$. 
Again ignoring any prefactors the parametric dependence of such 
quadrapole power loss is 
\begin{equation}
P_{\rm gravity} \sim {G_D M^2\over r_h^{D-2}}\ .
\end{equation}
The ratio of gauge to gravitation radiation in the balding phase
is then parametrically 
\begin{equation}
{P_{\rm gauge}\over P_{\rm gravity}} \sim {\alpha \over (r_h M_p)^{D-2} }\ 
\end{equation}
suggesting that gravitational radiation dominates.  
It is intuitively apparent that 
with order one gauge charges, gauge radiation is insignificant
in the large mass limit.
 
In four dimensions the total mass loss by classical gravitational radiation 
in the balding phase for an excited black hole produced by 
collision of neutral relativistic 
particles is estimated to be 16\% \cite{D'Eath:1993gr}.
This result should be indicative of the total energy lost
by an excited black hole during the balding phase  
also in the higher dimensional case 
since first, gravitational radiation is expected to dominate, 
and second because the energy-momentum multipole moments 
generated during the process of formation take values within the 
Standard Model brane. 
For production of large mass excited black holes by collision 
of relativistic charged particles in the high energy limit, 
numerical work at the classical level could
significantly improve these rough estimates.
For black hole masses near the fundamental Planck scale 
these estimates may receive potentially important 
quantum corrections.

The gauge charges inherited by the black hole from the initial 
state partons should discharge through the emission of a small 
number of quanta 
via the Schwinger process \cite{Zaumen,Gibbons:1975kk}.
This should take place either during the balding phase, or 
near the beginning of the evaporation phases discussed below. 

So at the end of the transient balding phase, an excited black 
hole produced in a high energy collision has lost most 
of its hair and is characterized 
essentially by the mass and angular momentum, 
and is therefore described by a spinning Kerr solution.


\sect{Black Hole Spin-Down and Evaporation}
\label{sec:evap}

After the balding phase, the black hole will decay 
more slowly via the semi-classical 
Hawking evaporation process\cite{Hawking:1975sw}.  It emits modes
both along the brane and into the extra dimensions, 
as illustrated in Fig. 2.  If the Standard Model
is comprised solely of brane modes, the bulk modes will be gravitational
and thus invisible.  
Furthermore, as argued in  \cite{Emparan:2000rs},
radiation along the brane is the dominant mechanism for mass loss. 
This follows from the observation that Hawking evaporation 
takes place predominantly in the $S$-wave. 
The emissivity to a given brane or bulk mode is then roughly 
comparable, and the large number of Standard Model brane modes
then dominate the evaporation process. 
Our discussion therefore neglects the bulk modes.  
We also assume that the only
relevant modes on the brane are those of the Standard Model, although the
discussion easily generalizes.  (Note in particular that in  
four dimensions for a large number of
scalar modes, there is no Schwarzschild phase discussed below; 
$J/M^2$ asymptotes to a
fixed value\cite{Chambers:1997ai}.)

\begin{figure}[ht]
\begin{center}
\epsfxsize= 3.95 in
\leavevmode
\epsfbox[0 0 577 259]{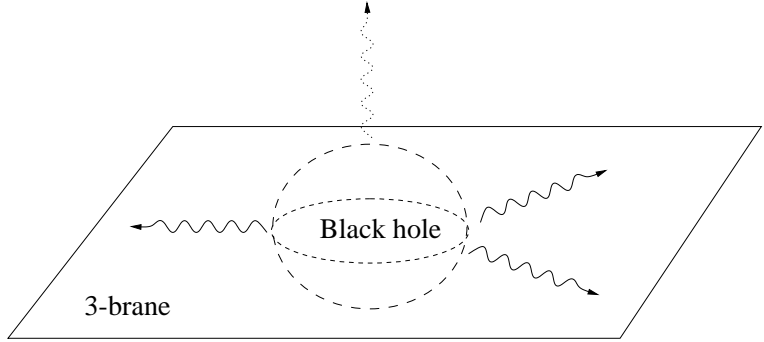}
\end{center}
\caption[f2]{A $D$-dimensional black hole bound to a 3-brane.  The black hole
emits Hawking radiation predominantly into brane modes (solid lines) and 
also into bulk modes
(dotted lines).  Grey body factors for brane modes are determined from the
metric induced by the $D$-dimensional black hole geometry on the brane.}
\label{bhevap}
\end{figure}

For black holes with temperatures down to of order 100 GeV,
all Standard Model
particles may be treated as essentially 
massless; for temperatures smaller than 
this phase space suppression for the heavy gauge bosons and top quark must 
be included. 
For a massless particle the emission rate per unit
energy $E$ and time is
\begin{equation}
{dN_{i,E,\ell,m,\lambda}\over dE dt} ={1\over 2\pi}
{\gamma_{i,E,\ell,m,\lambda} \over
\exp\left\{(E-m\Omega)/T_H\right\} \mp 1}
 \label{emrate}
\end{equation}
where here $i$ denotes species, $\ell,m$ angular quantum numbers, $\lambda$
polarization, and
\begin{equation}
\Omega= {1\over r_h} {a_* \over 1+a_*^2}\equiv {1\over r_h} \Omega_*
\end{equation}
is the surface angular frequency of the black hole, and 
$\Omega_*$ the dimensionless frequency. 
The $\gamma$ are gray-body (tunneling) factors which 
modify the spectrum of emitted particles from that of a perfect
thermal black body. 
Estimates for grey body factors in the four dimensional case 
are given in \cite{Page:1976df,Dewitt:1975ys}.
The grey body factors for Hawking emission into brane modes
in the higher dimensional case will differ quantitatively 
from the purely four dimensional case, but should have the 
same qualitative features. 
For $E r_h\ll 1$ they
vanish as $E^{2 \ell+1}$,
whereas for $E r_h\gg 1$ geometric optics predicts the $J=0$ values
$\gamma\approx \theta(K r_h E -\ell)$, with $\theta$ the step function
and $K$ a constant, and the $J\neq0$ values
$\gamma\approx e^{-\zeta \ell}$
with $\zeta = {\cal O}(1)$.

The evaporative time evolution of $M$ and $J$ for the black hole 
follows by summing (\ref{emrate})
over modes to which the black hole can evaporate.  The relevant
equations are most easily studied in dimensionless
variables\cite{Page:1976ki};
in
particular, define $x=E r_h$ and $T=T_*/r_h$.  The evolution equations
then become
\begin{equation}
Mr_h^2 {d\ln M\over dt} = -f\; ;\; {Jr_h\over a_*} {d\ln J\over dt} = -g
\label{MJeqs}
\end{equation}
with
\begin{equation}
\left(\matrix{f\cr g}\right) =  {1\over 2\pi} \sum_{i,\ell,m,\lambda} 
  \int_0^\infty
dx {\gamma_{i,E,\ell,m,\lambda}(x,a_*)\over
\exp\left\{ (x-m\Omega_*) /T_*\right\} \mp1}
\left(\matrix{x\cr m/a_* }\right)\ .
\label{decayrates}
\end{equation}
Dimension dependence enters these equations both through the 
grey body factors and the dependence of the Schwarzschild radius
on mass and angular momentum.

The parametric dependence of the black hole lifetime on mass follows 
directly  from (\ref{MJeqs})  
\beq
\tau = { C \over M_p} \left( {M \over M_p} \right)^{(D-1)/(D-3)} 
\eq
where $C$ is a numerical constant obtained by integrating 
(\ref{MJeqs}), and which depends on 
the dimensionality $D$, and initial angular momentum $J$. 
This parametric dependence may also be 
derived from the Stefan-Boltzmann law associated with Hawking 
emission. 
Equations (\ref{MJeqs}) have been solved numerically
in the four dimensional 
case by Page\cite{Page:1976ki}. The half life for spin-down is
computed from
\begin{equation}
{d\ln J\over d\ln M} = {D-2\over 2} {g\over f}\ .
\end{equation}
In the four dimensional case this half-life is
$7\%$ of the black hole lifetime. 
In the present case $f$ and $g$ differ from Page's
in the dimension dependence of the 
grey body factors and in the number of species
considered, but the results should be similar.  During spin-down the
angular momentum is shed in quanta with typically $\ell=m\sim 1$ and energy
$E \sim 1/ r_h$.  In the four dimensional case 
this phase accounts for about $25\%$ of the mass
loss.
The remaining mass is lost in 
a Schwarzschild phase characterized by 
the decay of the $J\approx0$ black hole.
In the four dimensional case, the constant $C$ is numerically found to 
be\cite{Page:1976df,Sanchez:1978vz,Kapusta:1999yy}
\begin{equation}
C^{-1} \simeq \left( 40 n_0 + 19 n_{1/2} + 7.9 n_1 +0.90 n_2\right) \times
10^{-3} 
\label{Ceq}
\end{equation}
where $n_s$ denotes the total number of polarization degrees of freedom 
for spin $s$. 
For the Standard Model, this gives $C \simeq 0.5$ in four dimensions.  
If the dimension dependence of the grey body factors is ignored, 
this may be used to crudely estimate the constant $C$ in the
higher dimension case by taking account of the dimension 
dependence of the Schwarzschild radius on mass. 
In $D=10$ this gives roughly $C \sim 6.5$.

The energy spectrum of particles emitted in the Hawking process
is derived by integrating (\ref{emrate}) over the lifetime of
the spin-down and Schwarzschild phases up to a cutoff time $\tau_P$
where the Planck phase discussed below begins. This gives
\begin{equation}
{dN_{i,E,\ell,m,\lambda}\over dE} = \int_0^{\tau - \tau_p}
  {dt\over 2\pi}
{\gamma_{i,E,\ell,m,\lambda}(r_h(t) E ,a_*(t))
\over
\exp\left\{\left[r_h(t)E-m
\Omega_*(t)\right]/T_*(t)\right\} \mp 1}\ .
\end{equation}
In particular, consider the Schwarzschild phase.
For $E\gg1/r_h \sim T_H$, the scaling of the grey body factors discussed  
above gives 
$\sum_{l,m}\gamma \propto (r_h E)^2$. 
This results in \cite{MacGibbon:1991tj}
in a spectrum which falls like $E^{1-D}$ at high energies, and 
is power suppressed \cite{Page:1976df} at low energies
\beq
{dN \over dE} \sim \left\{ 
\begin{array}{lccr} {E^{1-D}} & &:& ~~ E \gsim T_H \\
  E^{\eta} & &:& ~~ E \lsim T_H \\
\end{array}
\right.
\eq
where $\eta$ is a positive power,
and $T_H$ is the Hawking temperature
of the initial black hole. 
The dominant radiation is at energies 
$1/r_h \sim T_H$ determined by the initial black-hole mass.  
The total number of particles emitted is characteristic 
of the entropy
(\ref{Sbh}) of the initial black hole 
$N \sim S_{bh}$.
The evaporation phases may therefore be thought of as literally 
the evaporate escape of the degrees of freedom which make up the  
black hole state produced in the collision. 
Numerical estimates for $T_H$ and $S_{bh}$ in $D=10$ 
are presented in the next section. 
Numerical study of the evaporation equations is required to
improve all the rough estimates presented above. 

The relative emissivities for various types of particles
depend on the grey body factors. 
In four dimension these ratios may be extracted from the relative 
coefficients given in (\ref{Ceq}).
Summing over four-dimensional
transverse degrees of freedom, the relative emissivity 
for massless spin 0, spin ${1 \over 2}$, spin 1, and spin 2 particles 
in the four dimensional case is 
42\%:40\%:16\%:2\%.
These ratios should be indicative of those in higher dimensions. 
Numerical calculation of grey body factors,  
$\gamma_{i,E,\ell,m,\lambda}$, in the higher dimensional case should 
be pursued in order to improve these estimates. 

Note that for black holes with $r_h> R_c$, the above
expressions and estimates for the lifetime, decay spectra, {\it etc.}
must be modified to account for the effective change in dimensionality
discussed in section \ref{sec:grav}.

A discussion of the end point of Hawking evaporation requires
a full theory of quantum gravity. 
The semi-classical description breaks down when the Hawking 
temperature reaches the fundamental scale $T_H \sim M_p$. 
At this point the black hole reaches a final Planck phase 
of decay. 
We expect the black hole in this phase to decay 
to several quanta with energies of order the Planck or string
scale. 

Finally, in the case of 
primordial black holes in four dimensions with 
the standard Planck scale, there are
controversial
claims that a photosphere forms around an evaporating black hole, 
and thermally degrades the Hawking spectra \cite{Heckler:1997qq}.  
However, in the higher dimensional case with TeV Planck scale, 
the much smaller total entropy of a given mass black hole implies
that the outgoing Hawking radiation is sufficiently dilute 
so as not to thermalize. 
This may also be understood from the much shorter lifetime in the 
higher-dimensional case which 
implies that the outgoing radiation is emitted in a relatively 
thin shell of thickness order the lifetime $\tau$, 
which is too thin to thermalize.


\sect{Experimental Signatures}
\label{sec:sig}

The potentially large production cross section at 
hadron colliders presents the possibility 
of studying black hole formation and decay in some detail. 
The various stages of 
decay described above give rise to distinctive 
distributions of decay products in both type, energy spectrum, 
multiplicity, 
and angular distribution. 
For definiteness we consider the signatures which could 
be observed at the LHC 
associated with parameters of $M_p \simeq 1$ TeV and $D=10$.
The signatures for other space-time dimensions are qualitatively 
similar, and modifications for $M_p \gsim 1$ TeV are mentioned  
briefly below. 

At the LHC, because of the rapidly falling parton structure
functions a typical black hole of any given mass is produced without
a large boost in the laboratory frame, 
$\langle \gamma \beta \rangle \lsim 1$. 
This implies that the decay products are also not highly 
boosted in the laboratory frame. 
So the angular and energy distributions described below are largely 
preserved in the laboratory frame for a typical 
event, although there are exceptional events with sizeable boost factor
for lower mass black holes. 

The first stage of decay for an excited black hole produced in a 
high energy collision is  the balding phase. 
As discussed in section \ref{sec:production}
the energy lost 
in the 
process of shedding multipole hair is
a small but non-negligible
fraction of the excited black hole mass, perhaps 15\%. 
In the large mass
limit, gravitational radiation is expected to dominate the energy
loss in this phase. 
For black hole masses not too far above the fundamental Planck 
scale some fraction of the energy may be emitted in gauge 
quanta. 
The energy of the quanta emitted in this phase is determined 
by the multipole frequencies which are characterized by the 
Schwarzschild radius, $E \sim 1/r_h$. 
This energy scale coincides with that of the Hawking temperature
at the end of the balding phase. 
As discussed below, the Hawking temperature of a 10 TeV black 
hole for $D=10$ is roughly 150 GeV, and slightly less
for a smaller number of space-time dimensions.  
In this case the balding phase could give rise to 
probably at most only a few
gauge quanta, predominantly gluons, with energies of the 
order of 100-200 GeV. 
Distinguishing any visible quanta emitted during the balding phase 
from those emitted during the evaporation phase discussed below 
would seem to be difficult. 

The highest multiplicity of particles from black hole decay 
comes from the 
spin-down and Schwarzschild Hawking evaporation phases. 
The characteristic energy scale for these particles 
is the initial Hawking temperature of the black hole
after the balding phase. 
Numerically, the Hawking temperature for $D=10$ is 
\beq
T_H \simeq 0.2 M_p~(M_p/M)^{1/7}\ .
\eq
The distribution of energies 
extends up to roughly the 
fundamental Planck scale with a spectrum 
$dN / dE \sim E^{1-D}$. 
The total number of particles emitted by evaporation in this phase 
is roughly the entropy of the initial black hole after the 
balding phase. 
Numerically, the entropy for $D=10$ is  
\beq
S = {7M \over 8T_H} \simeq 4~(M / M_P)^{8/7}\ ,
\eq
so a large number of relatively hard primary partons arise 
from the evaporation phase. 
For example, for a 10 TeV mass black hole, of order 50  
quanta with a typical energy of order 150-200 GeV
result from the evaporation phase. 
As described below, almost all of these emitted particles
appear as visible energy. 
For smaller space-time dimension the Hawking temperature
is slightly lower for a given black hole mass. 
In this case a slightly higher multiplicity of somewhat lower
energy quanta would be released in the evaporation phases. 

Perhaps the most distinctive feature of the evaporation phase, 
aside from the high multiplicity, is the distribution  
in type of particle emitted. 
Because of the large fraction of 
Standard Model states which are strongly interacting,
most of the emitted particles are strongly interacting. 
As described in section \ref{sec:evap}, the relative emissivities
depend on the intrinsic spin of the emitted particle through
the grey body factors. 
Using the relative emissivities quoted in section \ref{sec:evap},
the Standard Model fractions of 
quarks and gluons, leptons, massive gauge bosons,
neutrinos and gravitons, Higgs boson, 
and photons emitted from a non-rotating black hole 
in four dimensions, ignoring 
particle masses, is   
72\%:11\%:8\%:6\%:2\%:1\%.  
Accounting for decay of top quarks, massive gauge bosons, and the Higgs
boson, the ratio of hadronic to leptonic activity (primary 
$e$, $\mu$, and $\tau$) in the evaporation 
process in this case is roughly 5:1, while the ratio of hadronic to photonic 
activity is roughly 100:1. 
Taking account of heavy quark and tau semi-leptonic 
decays would decrease these ratios slightly. 
The specific ratios in the higher dimensional case with Hawking 
evaporation along the Standard Model brane requires integration 
of the evaporation equations given in section \ref{sec:evap} 
including appropriate grey body factors and black hole angular
momentum. 
However, the four dimensional values are expected to be
indicative of those for the higher dimensional case.
The fraction of energy which is visible 
resulting from the evaporation phases 
is therefore expected to be in the 85--90\% range. 

Additional states to which the black hole could
evaporate, such as supersymmetric partners, would of course 
modify the specific ratios of final state partons.  
However, such states which decay to visible particles 
are sure to be identified at the LHC, 
and so the total amount of Hawking radiation which appears
as visible energy, as well as the 
ratios of hadronic to leptonic and hadronic to photonic 
activity 
will be calculable parameters in the large black
hole mass limit. 

Another feature of the evaporation phase is the angular 
distribution of emitted particles. 
As described in section \ref{sec:production},
a typical black hole is produced with a large angular 
momentum. 
The spin-down process of Hawking evaporation emits quanta predominantly  
in the $\ell=m \sim 1$ modes. 
The angular dependence of the particles emitted during the 
evaporation phase is then roughly 
\beq
{dN \over d \varphi}  \sim N_0 + 
      2 N_1 ~\sin^2 \varphi
\eq
where $N_0$ and $ N_1$ are the number of particles
radiated in the Schwarzschild and spin-down evaporation processes
respectively where the subscript refers to the $\ell$ value, 
$N_0+ N_1 \simeq N $, and 
$\varphi$ is the angle
with respect to the spin axis in the rest frame. 
In four dimensions the spin down phase accounts for about
25\% of the evaporative mass loss, and is expected to be similar
in the higher dimensional case with Hawking radiation on the brane. 
It might therefore be possible in large multiplicity events 
to discern the magnitude of initial black hole spin and 
direction of the spin axis from the 
angular distribution of emitted particles. 
The distribution of both the magnitudes and spin axis directions 
measured from a large number of events 
would provide information about the non-perturbative 
process of black hole formation which determines the initial spin. 
For example, deviations of the spin axis from a direction perpendicular
to the collision axis in the rest frame may occur 
for black hole masses not too far above the fundamental Planck scale.  


The endpoint of the black hole evaporation is the Planck phase 
in which the black hole completely decays. 
Without a fundamental theory of gravity it is hard to quantify 
this phase. 
But any visible partons emitted in this phase would 
have energies characteristic of the Planck or perhaps 
string scales. 
The identity and distributions 
of the highest energy final state partons within 
a black hole event would provide information about the Planck phase
and end point of Hawking evaporation. 

Because of the large cross section, 
large total visible energy, and 
large multiplicity of final state partons,  
black hole production gives rise to very spectacular events
at a hadron collider.  
For a 10 TeV black hole with $D=10$, on the order of 50 visible final 
state primary partons result, each with typical energy 
in the 100-200 GeV range from the balding and evaporative decay 
phases, with a few hard visible partons up to energies 
of order the fundamental Planck scale from the 
end of the evaporation phase and Planck decay phase. 
Since most of the black hole decay products result from the 
evaporation phase, 
the ratio of the total hadronic to leptonic activity is expected to 
be roughly 5:1. 
In addition, most of these particles are emitted in    
$\ell,m \lsim 1$ modes leading to a fairly spherical 
distribution of primary final state partons 
in the black hole rest frame. 
Since a typical black hole is produced with only a moderate boost factor, 
this results in events with a high degree of sphericity.  
For a completely spherical event corresponding to a
spinless black hole at rest in the lab frame, the 
transverse energy is ${1 \over 2}$ of the total energy. 
The moderate boost and high sphericity imply that the total 
visible transverse energy of a typical black hole event
is between ${1\over 3}$ and ${1 \over 2}$ of the total 
deposited visible energy. 

Another feature of the large multiplicity in a black hole event 
is that any missing energy either from primary 
emission by the black hole of gravitons, neutrinos, or other 
non-interacting particle such as a (quasi)-stable
neutralino,
or from neutrinos in subsequent cascade decays tends to 
average out within a given event. 
However, there can be exceptional events in which the missing 
energy fluctuates upward, from for example, 
emission  of two charged or strongly interacting particles 
recoiling against a hard graviton in the Planck decay phase. 

Since the transverse energy is an invariant it provides a 
very good measure of the black hole mass. 
This is true on a statisical basis if the initial black hole
spins are averaged, and is also true for a given large 
multiplicity event if the 
relative multiplicities $N_0$, $N_1$ discussed above can 
be extracted. 
As discussed above, most of the energy emitted in the 
evaporation phases is visible. 
So depending on the spin, the black hole mass should 
therefore by very roughly 
of order twice the visible transverse energy -- this 
ratio could be reliably calculated by numerical simulation 
of cascade decays of the primary partons to determine 
the precise fraction of energy which is visible. 

Special 
purpose triggers are not required to accept black hole events
because of the large total transverse 
hadronic energy and non-negligible leptonic fraction.
Even without a dedicated search, such events would appear
in any number of new physics analysis which utilize 
hadronic or leptonic activity. 
In fact, if the fundamental Planck scale is a TeV, because
of the relatively large cross section, it is
likely that black hole production and decay would represent
a significant background in many new physics searches. 

The most striking features of black hole decay are both the large
multiplicity and total transverse energy of the decay products. 
At a hadron collider an obvious cut to select black hole 
events is therefore 
large total transverse hadronic energy of at least 
a few times the fundamental Planck scale, and (very) large multiplicity 
of relatively hard jets. 
A requirement of relatively hard 
leptonic activity could also be applied.  
The requirement of a large multiplicity of final state
partons significantly reduces the background from perturbative
processes which typically only have a few hard final state 
partons from cascade decays of the primaries. 
In addition, black hole events have the feature that 
the multiplicity and average final state 
primary parton energy in any event is correlated 
with the black hole mass (as indicated by the 
event total transverse visible energy) in a manner determined by the 
Hawking evaporation process. 
Specifically, the multiplicity is higher and average energy 
per primary final state parton lower for higher mass events.
This is another manifestation of the infrared--ultraviolet 
connection of gravity -- higher energy events have lower
energy per particle. 
A growing parton--parton cross section 
for events satisfying these cuts, along with 
a roughly 5:1 ratio of hadronic to leptonic
activity would represent a smoking gun for black hole production 
and decay.

A calorimetric measurement of the number of identified 
black hole events 
as well as the average multiplicity and final state parton 
energy as a function of total event energy would give a measure
of the black hole production cross section. 
This in turn is sensitive the dimensionality and geometry 
of the extra dimensions. 

A measurement of the distribution of 
multiplicities and spectra of primary final state partons energies 
as a function of black hole mass, as indicated by 
event transverse visible energy, over a large number of events would 
allow a quantitative test of the Hawking evaporation process. 
Even though black holes produced in high energy collisions
Hawking radiate mainly to Standard Model particles on the brane, 
a precise measurement of the decay spectrum would be 
sensitive to the number of extra dimensions through the effect 
on the 
evaporation evolution equations. 


The final signal of black hole production is the 
suppression of hard scattering processes at energies 
at which black hole production becomes important. 
As described in section \ref{sec:production}, 
perturbative hard scattering processes are Boltzmann
suppressed at energies well above the fundamental 
Planck scale by the statistical thermodynamic properties of black 
holes, or equivalently because such hard processes are hidden behind
the event horizon formed during collision. 
Such a suppression would be apparent in, for example, 
the Drell-Yan or two jet cross sections at very high energies. 

In summary, the spectacular experimental signatures associated 
with black hole production and decay at a hadron collider 
include 
%
%
\begin{itemize}
\item Very large total cross section with production rates at the 
LHC approaching up to of order 1 Hz possible. 
\item Parton level cross section {\it grows} with energy 
at a rate determined by the dimensionality and geometry of the 
extra dimensions. 
\item Large total deposited energy 
up to a sizeable fraction of the beam energy, with visible transverse
energy typically of order ${1 \over 3}$ the total energy. 
\item Large multiplicity events, with up to many dozens of relatively 
hard jets and leptons. 
\item Average energy per primary final state parton decreases 
with total event transverse energy as indicated by the relation 
between initial Hawking temperature and black hole mass.  
\item Ratio of hadronic to leptonic activity of roughly 5:1
from the Hawking evaporation phase.
\item High sphericity events due to large multiplicity and 
moderate black hole boost factor in laboratory frame. 
\item The angular distribution 
within a given event is characteristic of the initial black hole spin. 
\item Some events contain a few hard visible quanta 
with energy up to order the fundamental Planck scale from the 
Planck decay phase. 
\item Suppression of hard perturbative scattering processes 
at energies for which black hole production becomes important. 
\end{itemize}
A search for all of these features together should be essentially 
free of background from any perturbative physics or instrumental sources. 
An observation of these signatures would 
represent compelling evidence for black hole production and decay
and TeV scale gravity. 
It is likely that a detailed study of the potentially large 
number of such events could provide information about 
both the process of production of black holes in high energy collisions, 
as well as the Planck decay phase and end point of Hawking evaporation. 

The use of black hole signatures at the LHC to set a specific 
lower limit on 
the fundamental Planck scale would require additional work.  
This includes a detailed study of potential backgrounds. 
Just as important would be a theoretical estimate of the 
energy at which the black hole description of 
intermediate states in both production and 
decay becomes reliable,
which is not available at this time. 
However, the LHC center of mass beam energy is fixed at 14 TeV. 
And the applicability 
of the description of high energy scattering through 
black hole states is likely limited to energies 
more than at least a few times the fundamental Planck scale. 
So if the fundamental Planck scale is larger than a few TeV 
it seems unlikely that the non-perturbative effect of 
black hole production and decay 
is a relevant 
description of gravitational effects at the LHC -- perturbative 
gravitational effects would likely be more relevant.  
There is therefore a window of opportunity for $M_p \lsim$ few TeV
in which the LHC would be a black hole factory, with very dramatic 
signatures. 
The window is limited mainly by the center of mass beam energy, 
and the rapidly falling parton distributions. 
The window would of course be much larger for the VLHC.


\sect{Conclusions}

The signatures outlined above -- in particular very high multiplicity
events with a large fraction 
of the beam energy converted into transverse energy with a growing 
cross section 
-- should serve as clear signals
for formation and Hawking evaporation of 
black holes at colliders.  
The observability of black hole production and decay is, of
course, critically dependent on the magnitude of the fundamental Planck
scale $M_p$. 
But once this threshold is crossed, the production rate is
large and rapidly rising.  Note that for $D=10$ 
the present bound of about $M_p \gsim 800$ GeV \cite{missErev} 
from missing energy signatures due to perturbative graviton 
emission will not be significantly 
improved by the Tevatron Run II since this process is 
energy rather than rate limited by the rapidly growing density of 
perturbative graviton states.  We therefore 
arrive at the surprising conclusion that  
if the fundamental Planck scale is of the order of 
a TeV the LHC will be a black hole factory, and that this possibility cannot
obviously be excluded before LHC begins operation.
It is also amusing to note
that, if the Planck scale is indeed of order of a TeV, 
formation of black holes through binary collisions could be observed
at LHC by the end of the decade, quite possibly 
before being observed astrophysically by LIGO.

Perhaps one of the most stunning features of such a scenario is that, 
because of the infrared--ultraviolet
 properties of gravity, black hole production seems to
represent the {\it end} of experimental 
investigation of short-distance physics by relativistic 
high-energy collisions.
Through the
formation of event horizons, black hole formation 
in high center of mass energy scattering effectively cloaks 
hard processes.  
At high center of mass energies the non-perturbative production 
of black holes dominates all perturbative processes. 
And as we have argued, at high energies black
hole production is increasingly a long-distance, semi-classical process.
However, there {\it can} be interesting features in 
high energy scattering experiments as 
the thresholds $R_c$ for sizes or curvature scales of the extra-dimensions
are reached and passed.  
This would provide information about the structure of the
extra dimensions that is complimentary to that found by studying the
the perturbative graviton Kaluza-Klein spectrum
at lower center of mass energies. 
Note also that the growing cross sections for 
black hole production would simplify
design requirements for future colliders which typically anticipate a hard
scattering cross section which falls like the square of the center of mass
energy.

Collider study of black hole creation would certainly be an astounding
pursuit, although it may be that the most conceptually profound physics
would be unraveled at energies in the vicinity of the
Planck scale.  Here one would hope to reveal the microphysics of quantum
gravity and possible breakdown of spacetime structure.

\centerline{\bf Note Added}

While this work was in progress, we learned that some aspects of black
hole production were also under consideration by another group; that
work has appeared subsequent to our paper's appearance on the e-print
archive as \cite{Dimopoulos:2001hw}.

\subsection*{Acknowledgements}

We wish to thank T. Banks, L. Bildsten, D. Coyne,
J. Hartle, G. Horowitz, S. Hughes, J. Kapusta, J. MacGibbon, 
R. Myers, D. Page, 
A. Seiden, L. Susskind, and especially G. Horowitz for valuable
discussions, and T. Rizzo for discussions and for 
providing cross sections based on the CTEQ5 struction 
functions. 
This work was supported by National Science Foundation
grants PHY99-07949, and PHY98-70115, by the
DOE under contract DE-FG-03-91ER40618,
and by the Alfred P. Sloan Foundation.

\appendix
\section{Comparison of Conventions}

In this paper we normalize $M_p$ in a
convention useful in quoting experimental bounds\cite{Peskin:2000ti}.  
In these conventions,  the $D$-dimensional Newton
constant and the Planck mass are related by
\beq
M_p^{D-2} = {(2\pi)^{D-4}\over 4 \pi G_D}\ .
\eeq
At least two other conventions exist.  For example, bounds quoted in the
{\it Linear Collider physics resource book}\cite{Abe:2001nq} are quoted for
$M_D$ in
the convention of the first reference of \cite{missE}:
\beq
M_D^{D-2}= {(2\pi)^{D-4}\over 8 \pi G_D}\ .
\eeq
Thus
\beq
M_p = 2^{1\over D-2} M_D\ ,
\eeq
a small relative correction.

The paper by Dimopoulos and Landsberg \cite{Dimopoulos:2001hw} uses
somewhat different conventions,
\beq
M_{DL}^{D-2}= {1\over G_D}\ .
\eeq
Therefore the relation between the Planck masses in our two normalizations
is
\beq
M_{p}^{D-2}= 2^{D-6} \pi^{D-5} M_{DL}^{D-2}\ .
\eeq
In $D=6$ the difference is not great, $M_{p}= 1.3 M_{DL}$, but in $D=10$
the difference results in a substantial factor: $M_{p}= 2.9 M_{DL}$.


\end{document}